\documentclass[prl,showpacs,twocolumn,superscriptaddress]{revtex4}

\usepackage{graphicx}
\usepackage{amssymb}
\usepackage{color}
\usepackage{soul}

\sethlcolor{yellow}

\newcommand{\vvec}{\mathbf}

\newcommand{\del}{\partial}

\begin{document}

\title{Electrochemical response of biased nanoelectrodes in solution}

\author{Kentaro~Doi}
\email{doi@me.es.osaka-u.ac.jp}
\affiliation{Department of Mechanical Science and Bioengineering, Graduate School of Engineering Science, Osaka University, Osaka 560-8531, Japan}
\author{Makusu~Tsutsui}
\affiliation{The Institute of Scientific and Industrial Research, Osaka University, Osaka 567-0047, Japan}
\author{Takahito~Ohshiro}
\affiliation{The Institute of Scientific and Industrial Research, Osaka University, Osaka 567-0047, Japan}
\author{Chih-Chun Chien}
\affiliation{Theoretical Division, Los Alamos National Laboratory, Mail Stop B213, Los Alamos, New Mexico 87545, USA}
\author{Michael Zwolak}
\affiliation{Department of Physics, Oregon State University, Corvallis, Oregon 97331, USA}
\author{Masateru~Taniguchi}
\email{taniguti@sanken.osaka-u.ac.jp}
\affiliation{The Institute of Scientific and Industrial Research, Osaka University, Osaka 567-0047, Japan}
\author{Tomoji~Kawai}
\affiliation{The Institute of Scientific and Industrial Research, Osaka University, Osaka 567-0047, Japan}
\author{Satoyuki~Kawano}
\email{kawano@me.es.osaka-u.ac.jp}
\affiliation{Department of Mechanical Science and Bioengineering, Graduate School of Engineering Science, Osaka University, Osaka 560-8531, Japan}
\author{Massimiliano~Di~Ventra}
\email{diventra@ucsd.edu}
\affiliation{Department of Physics, University of California, San Diego, La Jolla, California 92093, USA}

\begin{abstract}
Novel approaches to DNA sequencing and detection require the measurement of electrical currents between metal probes immersed in ionic solution. Here, we  experimentally demonstrate that these systems maintain large background currents with a transient response
that decays very slowly in time and noise that increases with ionic concentration. Using a
non-equilibrium stochastic model, we obtain an analytical expression for the ionic current that shows these results are due to a fast
electrochemical reaction at the electrode surface followed by the slow formation of a diffusion layer. During the latter, ions translocate in the weak electric field generated after the initial rapid screening of the strong fields near the electrode surfaces. Our theoretical results are in very good agreement with experimental findings.
\end{abstract}


\maketitle

Single-molecule detection using micro/nano-fluidic devices has recently attracted quite a lot of attention~\cite{Zwolak2008,Branton2008,Howorka2009,Venkatesan2011}. In particular, solid-state nanopores~\cite{Storm2001,Fologea2005a,Dekker2007,Tsutsui2012}, in which translocated molecules can be detected by electrical signals~\cite{Zwolak2005,Lagerqvist2006,Lagerqvist2007}, are expected to become one of the core technologies for next-generation DNA and RNA sequencing.
As part of this approach, the tunneling current through nucleotides as they translocate through the nanochannel was recently measured using nanometer gap electrodes made by a mechanically controllable break junction (MCBJ)~\cite{Tsutsui2008a,Tsutsui2008b,Tsutsui2009,Tsutsui2010,Ohshiro2012}. In addition to information on single nucleotides, the measured electrical signals have contributions from  ionic currents due to the presence of the electrolyte solution.
Those signals are usually hidden within the inevitable noise in the nanofluidic system.
Clarifying the mechanisms responsible for the behavior of these ``background'' currents is critical to the development of DNA sequencing approaches based on electronic transport. Their study is also of relevance for any nanofluidic technology that employs electrodes in solution, e.g., for the improvement of the energy storage capabilities of electrochemical capacitors~\cite{Simon2008,Rica2012,Wei2012}.

In this letter, we experimentally investigate electrical currents generated by electrodes in an electrolyte solution. We find a transient electrical response that decays very slowly and an associated noise that increases with increasing ionic solution concentration. We develop a theoretical understanding of these effects that is in very good
agreement with the experimental observations and reveals the rich behavior contained in the electrochemical response.

\begin{table}
\begin{tabular}{c}
\includegraphics[width=1.05\linewidth]{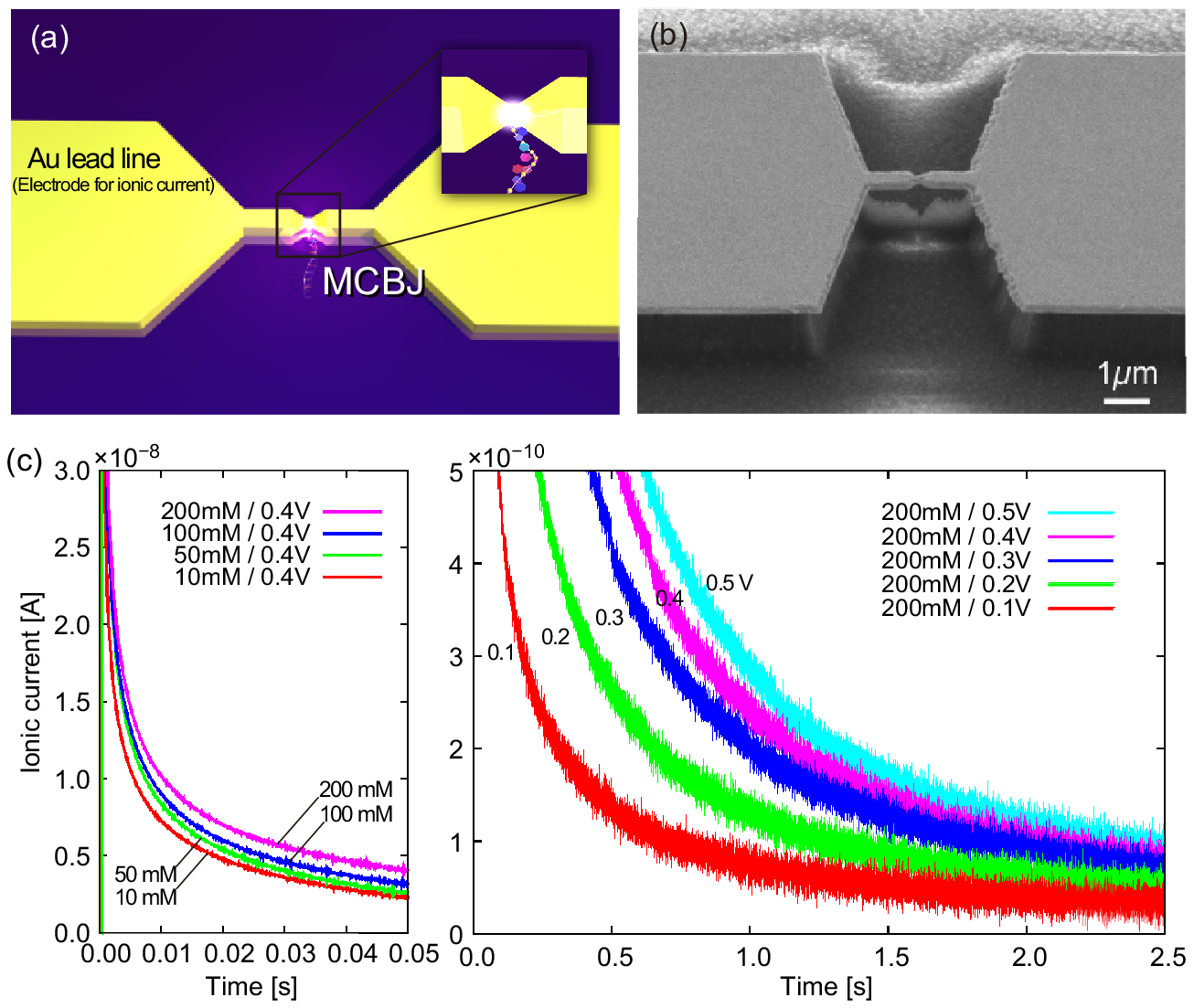} \\
\multicolumn{1}{p{246pt}}{FIG. 1. (Color online) (a) Schematic illustration and (b) SEM image of MCBJ electrodes, (c) transient current response of NaCl aqueous solution depending on salt concentrations and applied electric potentials, and (d) current response for a concentration of 200 mM NaCl at various voltages. Slow current decays of about 1 s can be observed after rapid increase in the ionic current (see also supplementary Fig. S1 and Table S1).}
\end{tabular}
\end{table}
Figures 1(a) and 1(b) show a schematic illustration and scanning electron microscopy (SEM) image, respectively, of a MCBJ made with Au electrodes. The lithographically defined nano-junction is fabricated on a polyimide-coated phosphor bronze substrate.
It consists of 30 nm thick Au/Cr wires of approximately 10 mm end-to-end length and width tapered from 10 $\mu$m to 1$\mu$m along the length.
We create a pair of nanoelectrodes using a self-breaking technique~\cite{Tsutsui2008a}, which forms 1 nm electrode gap with atomically sharp Au tips sensitive enough to detect single molecules by a tunneling current~\cite{Tsutsui2009,Tsutsui2010,Ohshiro2012}.
In our experimental setup, the electrodes are symmetrically placed in a liquid container of 2 mm diameter.
Therefore, large surfaces ($\sim10^{-8}$ m$^2$) of the lead line are exposed to the electrolyte solution, which cause the background current with noises as shown in Fig. 1(c).
Here, the electrical response of NaCl solution was measured for various salt concentrations under constant dc voltages at 10 kHz by using a custom-built setup~\cite{Ohshiro2012} (see also supplementary materials).
From Figs. 1(c) and 1(d), the transient response includes a rapid increase and subsequent slow decay of the ionic current.
Particularly, the time constant increases with increasing molarity and applied potential, and is on the order of 1 s for the parameter regimes studied (see also supplementary Fig. S1 and Table S1).
Assuming an equivalent circuit, this system would then consist of a 1 G$\Omega$ resistance evaluated from the current--voltage characteristics and a 1 nF capacitor.
However, such a huge resistance and small capacitance in the narrow space cannot be explained by the conventional macroscopic models. We then set to understand these features by employing a microscopic model.

Ionic motion in aqueous solutions can be expressed by Newton's equation of motion with fluctuation and dissipation.
The motion of a particle affected by collisions from solvent molecules in external force fields is expressed by
$m\dot{v} = -{\xi}v + F + R$,
where $m$ is the mass, $v$ is the velocity, $\xi$ is the friction coefficient, $F$ is the force caused by the potential field, and $R$ is the random force from the solvent molecules.
In such a system, when an electric field is applied, the uniform field spreads in the solution at first and ions will respond at each point in space.
Subsequently the electric field rapidly shrinks due to the screening effect very near the electrode surfaces~\cite{Kimball1940,Hamelin1996a,Schoch2008}.
In the next stage, the ions gradually form a diffusion layer, interacting with each other.

In order to elucidate the scale of phenomena (and hence a better understanding of the underlying physics) we consider the one-dimensional ionic current densities $j_i(x,t)$ due to the migration and diffusion of each of the ionic species $i$ in aqueous solution.
Taking into account the stochastic process in electrolyte solutions, ionic motions can be expressed by a Nernst--Planck equation for each species~\cite{Schoch2008}:
\begin{eqnarray}
\label{NP}
\frac{\del \rho_i}{\del t}&=&-\frac{\del j_i}{\del x};~ j_i(x,t)=\frac{F_i}{\xi_i}\rho_i-D_i\frac{\del \rho_i}{\del x},
\end{eqnarray}
where $\rho_i(x,t)$ is the charge density expressed by the valence $z_i$, the number density $n_i(x,t)$, and the elementary charge $e$: $\rho_i=z_ien_i$, and $D_i$ is the diffusion coefficient.
The coordinate $x$ is defined in an interval of $x\in(0,L)$ where the electrode surface is at $x=0$ and the thickness of diffusion layer is $L$.
Ionic migration is driven by the electrostatic force due to the applied potential and thus $F_i$ can be represented by the gradient of potential $\phi$ such that $F_i=-z_ie\nabla\phi$.
The relation between $\phi$ and $\rho_i$ is expressed by the Poisson equation: $-\varepsilon\triangle\phi=\sum_i \rho_i,$
where $\varepsilon$ is the dielectric constant of the solution.
The summation is taken for all species that contribute to $\phi$.
The ionic density and the electrostatic potential should be determined self-consistently to solve the non-linear partial differential equation.
Taking into account the effect of noise as shown in the experimental result (Fig. 1(c)), source terms are added and thus Eqs. (\ref{NP}) become
\begin{eqnarray}
\label{Eq2}
\frac{\del \rho_i}{\del t}&-&\frac{z_ie}{\xi_i}\frac{\del}{\del x}\left(\frac{{\rm d} \phi}{{\rm d} x}\rho_i\right)-D_i\frac{\del^2 \rho_i}{\del x^2} \nonumber \\
&=&\sum_r f_{ir}\cos(\omega_{ir} t+\varphi_{ir})\delta(x),
\end{eqnarray}
where $f_{ir}$ are constants depending on the current density and noise, and $\varphi_{ir}$ are the phase shifts at $t=0$.
In the numerical solution, $\phi$ depends only on the displacement of ions and thus it is treated as independent of time in the short interval.
The smallest time step, which is large enough to represent the stochastic process, should be determined properly to maintain constraints at boundaries and electroneutrality.
The source terms represent noise generated at $x=0$ and induce external flux in the domain.
As a first step, we consider Gaussian white noise, although it is known that noises detected in micro/nano-fluidic devices usually show also flicker noise~\cite{Cossa2007,Smeets2008,Venkatesan2012}.
The parameters $\omega_{ir}$ and $\varphi_{ir}$ are generated by Gaussian~\cite{BM} and uniform probability distributions, respectively.

Rescaling $x$, $t$, and $\phi$ to make them dimensionless by
\begin{eqnarray}
x^*=\frac{x}{L},\ \ \ t^*=\frac{D_i}{L^2}t,\ \ \ \psi_i^*=\frac{z_ie\phi}{D_i\xi_i}=\frac{z_ie\phi}{k_{\rm B}T},
\end{eqnarray}
and, considering a unit surface, replacing $\rho_i$ by
\begin{equation}
n_i^*=\frac{L}{z_ie}\rho_i\exp\left[\frac{1}{2}\int^{x^*}\frac{{\rm d}\psi_i^*}{{\rm d}{x^*}'}{\rm d}{x^*}'\right],
\end{equation}
Eq. (\ref{Eq2}) becomes
\begin{eqnarray}
\label{Eq4}
& &\left[\frac{\del}{\del t^*}+\frac{1}{4}\left(\frac{{\rm d}\psi_i^*}{{\rm d}x^*}\right)^2-\frac{1}{2}\frac{{\rm d}^2\psi_i^*}{{\rm d}{x^*}^2}-\frac{\del^2}{\del{x^*}^2}\right]n_i^* \nonumber \\
& &=e^{\frac{1}{2}\int^{x^*}\frac{{\rm d}\psi_i^*}{{\rm d}{x^*}'}{\rm d}{x^*}'}\sum_{r} f_{ir}^*\cos\left(\omega_{ir}^*t^*+\varphi_{ir}\right)\delta(x^*),
\end{eqnarray}
where $f_{ir}^*={L^2f_{ir}}/{z_ieD_i}$ and $\omega_{ir}^*=L^2\omega_{ir}/D_i$.
For the other species, the equation can be derived in the same manner with a difference of factor $D_j/D_i$.
Here, $n_i^*$ and $\psi_i^*$ are expanded by the Fourier series:
$
n_i^*(x^*,t^*) = \sum_k c_{ik}(t^*) e^{ik\pi x^*}
$
and
$
\psi_i^*(x^*) = \sum_k g_{ik} e^{ik\pi x^*}
$
where $k=0,\pm 1,\pm 2, \cdots$.
$\delta(x^*)$ is also expanded similarly:
$
\delta(x^*) = \frac{1}{2\pi}\sum_k e^{ik\pi x^*}.
$
Each basis function is orthogonal on $x^*\in[-1,1]$.
In this model, a mirror symmetry is assumed at $x^*=0$ and 1.
Ions are adsorbed or reflected at $x^*=0$ and the concentrations correspond to those of bulk at $x^* \ge 1$, conserving the electro-neutrality in $L$.
Based on the description above, Eq. (\ref{Eq4}) becomes
\begin{eqnarray}
\label{Eq7}
\frac{{\rm d}\vvec{c}_i}{{\rm d}t^*}+\hat{H}_i\vvec{c}_i
=\frac{1}{2}\sum_{r}f_{ir}^*\cos\left(\omega_{ir}^*t^*+\varphi_{ir}\right),
\end{eqnarray}
where
\begin{eqnarray}
\label{Hkl}
H_{ikl}&=&\pi^2k^2\delta_{k,l}+\frac{\pi^2}{2}(k-l)^2g_{i(k-l)} \nonumber \\
      &-&\frac{\pi^2}{4}\sum_{k_1}k_1(l-k-k_1)g_{ik_1}g_{i(k-l+k_1)}.
\end{eqnarray}
$\psi_i^*$ are real functions and thus $\hat{H}_i$ are Hermitian matrices.
For the homogeneous case, we can solve Eq. (\ref{Eq7}) via its eigenvalues $\lambda_j$ and eigenvectors.
Using these solutions, we can also solve the inhomogeneous case.
If $\lambda_j$ are non-zero then
\begin{eqnarray}
c_{ik}(t^*)&=&\frac{1}{2}\sum_{j}\sum_{l}u_{kj}^{\dagger}u_{jl}\left[\frac{f_{ir=0}^*}{\lambda_j}(1-e^{-\lambda_jt^*})\cos\varphi_{ir=0}\right. \nonumber \\
&+&\left.\sum_{r\ne 0}f_{ir}^*I_{ir}^j(t^*)\right]
+\sum_{j}\sum_{l}u_{kj}^{\dagger}u_{jl}c_{il}^0e^{-\lambda_jt^*},
\end{eqnarray}
where $u_{kj}^\dagger$ are the conjugate transpose of $u_{jk}$, $c_{il}^0$ are constants determined at the initial condition and
\begin{eqnarray}
I_{ir}^j(t^*)&=&e^{-\lambda_jt^*}\int_0^{t^*}\cos\left(\omega_{ir}^*{t^*}'+\varphi_{ir}\right)e^{\lambda_j{t^*}'}{\rm d}{t^*}' .
\end{eqnarray}
At equilibrium, $\lambda_j$ should be zero because the concentration is in equilibrium.
As a consequence, $\rho_i$ are obtained and $\phi$ is given by solving the Poisson equation in the steady state (see supplementary materials).
Thus, we obtain
\begin{eqnarray}
\label{j2}
j_i(x,t) &=&\frac{2\pi z_ieD_i}{L^2}e^{-\frac{z_ie}{2k_{\rm B}T}\int^x\frac{{\rm d}\phi}{{\rm d}x'}{\rm d}x'}\sum_{k>0}k{\rm Im}\left[c_{ik}e^{ik\pi x/L}\right] \nonumber \\
&-&\frac{z_ieD_i}{2k_{\rm B}T}\frac{{\rm d}\phi}{{\rm d}x}\rho_i .
\end{eqnarray}

We apply this model to evaluate the ionic current near a Au electrode in a NaCl solution.
At the negatively charged cathode surface, Na$^+$ is highly concentrated regardless of anion species.
In the numerical analysis, $L$ is taken as a constant for each molarity.
The range of applied potentials is according to the practical potentials in aqueous solutions: $\phi(0)=-0.01$ to $-1$ V for $\phi(L)=0$ V fixed.
The valence is $+1$ and $-1$ for Na$^+$ and Cl$^-$, respectively.
The diffusion coefficients are known: $D_{{\rm Na}^+}$=1.35$\times$10$^{-9}$ m$^2$/s and $D_{{\rm Cl}^-}$=2.03$\times$10$^{-9}$ m$^2$/s~\cite{Kharkats1979}.
The dielectric constant and temperature are set to $\varepsilon$=78.4$\varepsilon_0$~\cite{Vidulich1967} where $\varepsilon_0$ is the dielectric constant of vacuum and $T$=298.15 K, respectively.
More details of computational methods are described in the supplementary materials.
As a first step, the time constant $\tau$=$L^2/D\lambda$ of the transient response can be evaluated by solving the homogeneous equation of Eq.~(\ref{Eq4}) for the two component system, focusing on the largest $\tau$ that mainly causes the slow decay of the response.
\begin{table}
\begin{tabular}{c}
\includegraphics[width=0.95\linewidth]{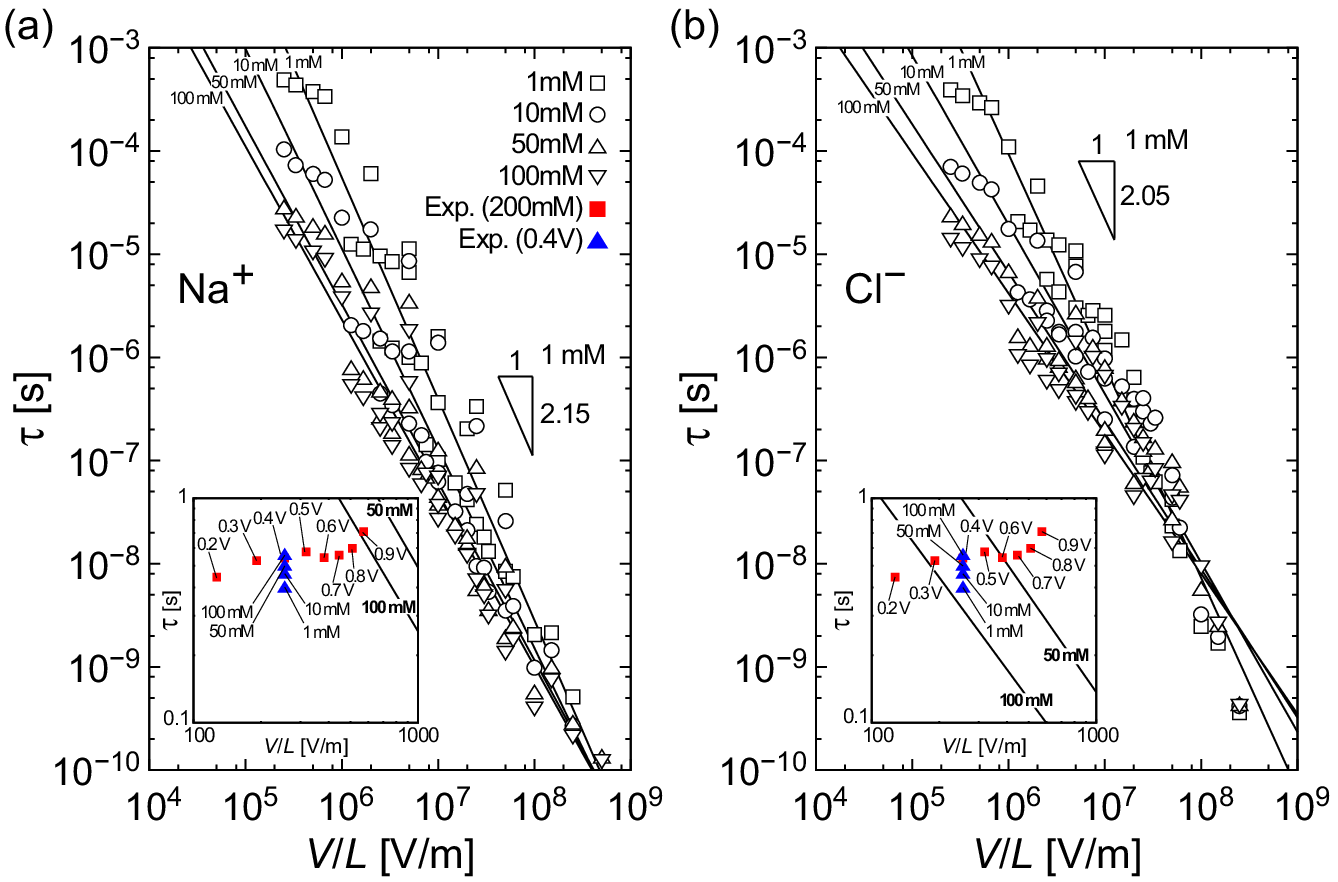} \\
\multicolumn{1}{p{246pt}}{FIG. 2. (Color online) Time constant $\tau$ of the transient response of (a) Na$^+$ and (b) Cl$^-$ in aqueous solution as a function of $V/L$ where $V=\phi(L)-\phi(0)$; $\tau$ and $V/L$ show $\tau\propto (V/L)^{-2.15}$ for Na$^+$ and $\tau\propto (V/L)^{-2.05}$ for Cl$^-$ fitting the computational data at 1 mM. Experimental results (Fig. 1(c)) are also shown in insets.}
\end{tabular}
\end{table}

Figure 2 shows $\tau$ for Na$^+$ and Cl$^-$ as a function of the ratio of $V=\phi(L)-\phi(0)$ to $L$ for the molarity of 1, 10, 50, and 100 mM, where $V/L$ merely expresses a fraction of the applied potential and the length, but not the electric field in the system.
It is found that $\tau$ tends to be proportional to $(V/L)^\zeta$ where $\zeta$=$-2.15$ for Na$^+$ and $-2.05$ for Cl$^-$ resulting from the fits of the computational data at 1 mM.
It can then be predicted that $\tau\propto(V/L)^{-2}$ at the dilute limit, since the second term on the left-hand side of Eq.~(\ref{Eq4}) becomes dominant as $n_i^*\sim0$.
On the other hand, the gradient decreases as the molarity increases due to the strong screening of the electrode surface.
These results suggest that ions rapidly respond to the strong electric field near the electrode surface and that on the other hand $\tau$ becomes large in the weak field that exists far away from the electrode.
The magnitude of $\tau$ is expected to be on the order of 1 s for the electric field of 10$^3$ V/m.
For comparison, $\tau$ evaluated from our experimental results (Fig. 1(c), supplementary Fig. S1, and Table S1) is also presented in Fig. 2.
In order to evaluate $\tau$ from the experimental results, the $L$ is fixed at 1.57 mm taking into account the longest length of the electric field line in the system (see Supplementary materials).
The theoretical evaluations are then in close agreement with the experimental results.
Furthermore, the series of $\tau$ resulting from the 200 mM solution is expected to show negative gradient if the $L$ becomes large with increasing the applied potential.

On the other hand, the above model cannot treat chemical reactions and the charge transfer explicitly.
In order to replicate constant currents at long time and noises with respect to the electrochemical reactions~\cite{Kimball1940}, we then apply source terms for Na$^+$.
\begin{table}
\begin{tabular}{c}
\includegraphics[width=0.9\linewidth]{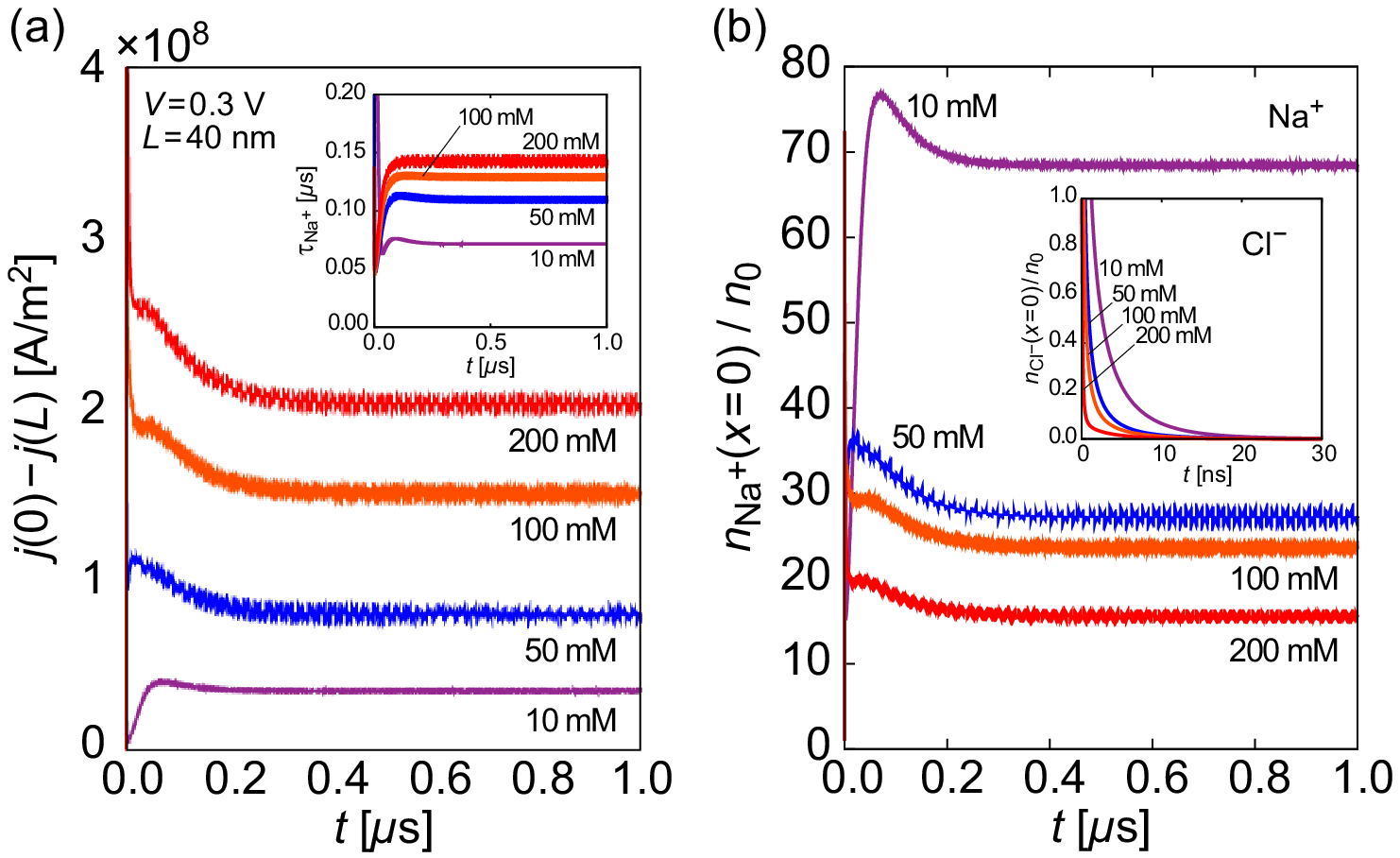} \\
\multicolumn{1}{p{246pt}}{FIG. 3. (Color online) Time evolution of (a) current density of 10 (purple), 50 (blue), 100 (orange), and 200 (red) mM NaCl solution where $\tau$ of Na$^+$ is also shown in inset, and (b) number density of Na$^+$ and Cl$^-$ (inset) at $x=0$, which are normalized by each bulk density $n_0$. Computations are performed for $V=\phi(L)-\phi(0)=0.3$ V and $L=40$ nm.}
\end{tabular}
\end{table}
Figure 3 shows current density and normalized number density of each species obtained from 10, 50, 100, and 200 mM NaCl solutions as a function of time.
Source terms --- to represent the Gaussian white noise --- are added to Eq.~(\ref{Eq2}).
The zeroth source term $f_{{\rm Na}^+r=0}$ is applied only for Na$^+$ in order to determine the current density at the steady state.
We set $f_{{\rm Na}^+r=0}$=$-\alpha e^2n_0D(\phi(L)-\phi(0))/Lk_{\rm B}T$, based on the Nernst--Einstein relation, where $n_0$ is the number density of bulk and $\alpha$ is a constant.
In addition, the amplitude of the noise is determined to be proportional to square root of the bulk density based on the surface charge density~\cite{Schoch2008} such that $f_{{\rm Na}^+r}=\beta\sqrt{n_0}$ with a constant $\beta$ common for each molarity.
$f_{{\rm Na}^+r}$ is determined from a fraction of $f_{{\rm Na}^+r=0}$ for 10 mM such that $\beta=-f_{{\rm Na}^+r=0}/\alpha\sqrt{n_0}$.
We perform computations for a case of $\alpha=1$, $\beta=100$, $L=40$ nm, and $V=0.3$ V.
The total ionic current density is evaluated as $\sum_i(j_i(x\to 0)-j_i(x\to L))$.
As shown in Fig. 3(a), rapid increases and subsequent decays in the net current density are observed at the moment when the electric potential is applied.
The time constant of Na$^+$ tends to become large as the molarity increases as shown in the inset of Fig. 3(a).
$\tau\sim0.1$ $\mu$s, obtained from $L=40$ nm and $V=0.3$ V ($V/L\sim 10^7$ V/m), closely corresponds to the result from Fig. 2 .
This trend is in close agreement with the experimental results (Fig. 1(c)).
Cl$^-$ also shows large $\tau$, but does not contribute to the current density at the cathode side due to the extremely low density near the surface.
As shown in Fig. 3(b), the response in the current density is obviously caused by the prominent increase and subsequent decrease of Na$^+$ at the surface.
Furthermore, the effect of noise tends to be relatively weaker as the molarity increases.
This is a reason why the noise is suppressed due to the highly screened surface.

\begin{table}
\begin{tabular}{c}
\includegraphics[width=0.97\linewidth]{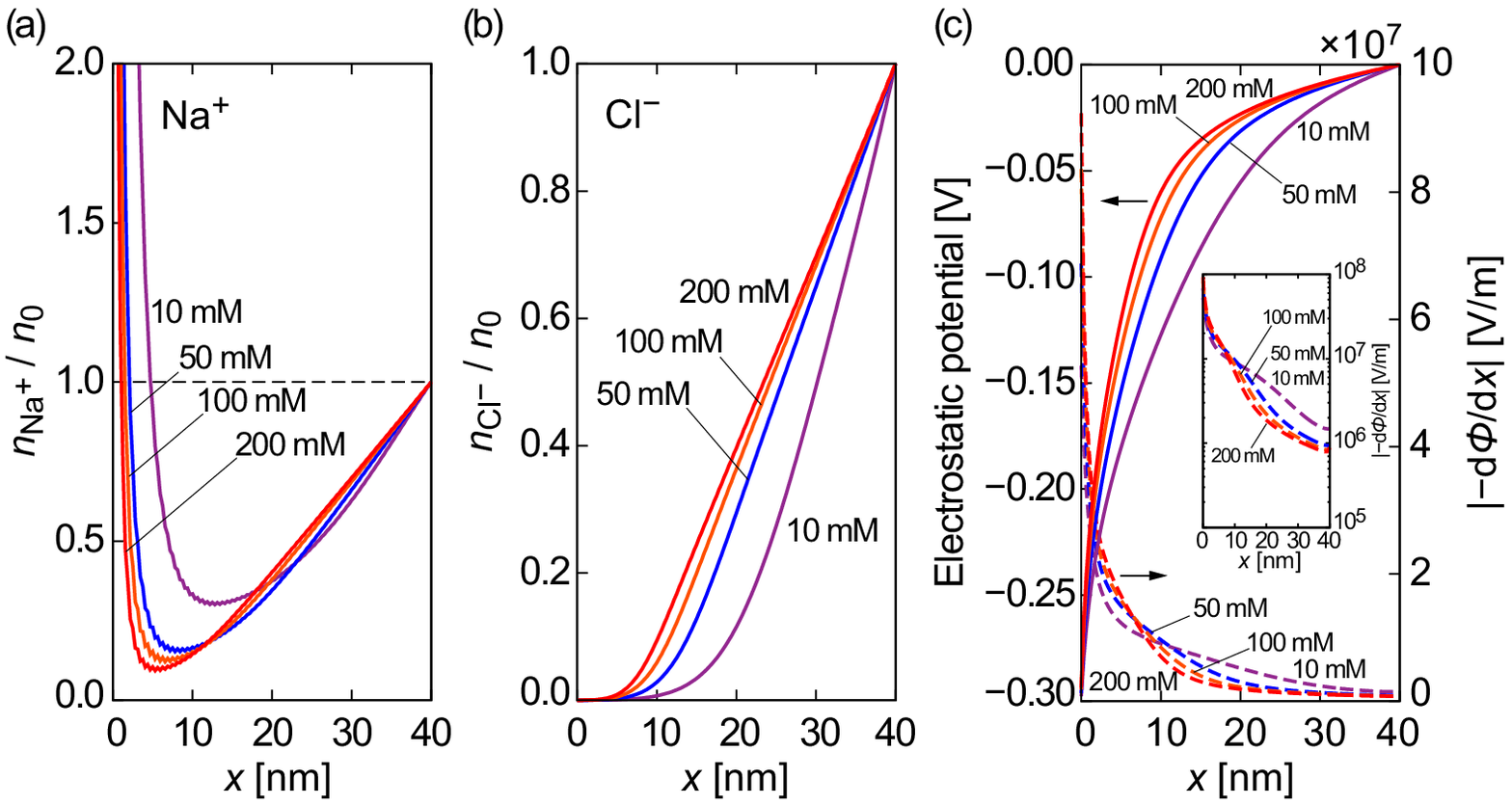} \\
\multicolumn{1}{p{246pt}}{FIG. 4. (Color online) Normalized density distribution of (a) Na$^+$ and (b) Cl$^-$ for 10 (purple), 50 (blue), 100 (orange), and 200 (red) mM and (c) the electrostatic potential (solid line) and the electric field (dashed line) that is also shown in logarithmic scale in inset. These characteristics are obtained from data at $t=1.0$ $\mu$s in Fig. 3 (see also supplementary Fig. S3).}
\end{tabular}
\end{table}
Figure 4(a) shows density distributions of Na$^+$ at $t=1.0$ $\mu$s.
For each concentration, a minimum peak of Na$^+$ is found near the surface.
This peak implies that part of the ions tends to adsorb on the electrode surface and the others separate forming density gradients.
In the case of Cl$^-$ as shown in Fig. 4(b), the concentration is depleted near the electrode surface and approaches its bulk density as $x$ increases.
Due to these distributions, the electrostatic potentials show extremely steep gradients near the electrode surface, as shown in Fig. 4(c).
It is found that the cathode surface is strongly screened by Na$^+$ and this tendency is apparent as the molarity increases.
In the 100 mM solution, the strength of the electric field is two orders of magnitude different between both ends.
Despite the strong screening, however, weak fields also seem to exist widely in the solution, which is inevitable to induce ionic current due to transport of electrolytes.
The above discussion can be applied to the anode side, although results may exhibit small differences due to the different diffusion coefficient of Na$^+$ and Cl$^-$.

In conclusion, we have reported a combined experimental and theoretical study of electrical currents of electrodes in solution. We have shown that this current is the result of different processes. In particular the long duration of the initial current response is due to the diffusion layer formation where ions respond to the weak electric fields remaining after the rapid screening at the electrode surface. These are relevant for DNA sequencing approaches presently pursued as well as electrochemical capacitors for high energy density storage.

This work was partly supported by the Japan Society for the Promotion of Science (JSPS) through its ``Funding Program for World-Leading Innovative R\&D on Science and Technology''.
CCC acknowledges the support of the U. S. DOE through the LANL/LDRD Program and MD partial support from NIH.

\bibliographystyle{apsrev}
\bibliography{refs}

\begin{thebibliography}{28}
\expandafter\ifx\csname natexlab\endcsname\relax\def\natexlab#1{#1}\fi
\expandafter\ifx\csname bibnamefont\endcsname\relax
  \def\bibnamefont#1{#1}\fi
\expandafter\ifx\csname bibfnamefont\endcsname\relax
  \def\bibfnamefont#1{#1}\fi
\expandafter\ifx\csname citenamefont\endcsname\relax
  \def\citenamefont#1{#1}\fi
\expandafter\ifx\csname url\endcsname\relax
  \def\url#1{\texttt{#1}}\fi
\expandafter\ifx\csname urlprefix\endcsname\relax\def\urlprefix{URL }\fi
\providecommand{\bibinfo}[2]{#2}
\providecommand{\eprint}[2][]{\url{#2}}

\bibitem[{\citenamefont{Zwolak and {Di Ventra}}(2008)}]{Zwolak2008}
\bibinfo{author}{\bibfnamefont{M.}~\bibnamefont{Zwolak}} \bibnamefont{and}
  \bibinfo{author}{\bibfnamefont{M.}~\bibnamefont{{Di Ventra}}},
  \bibinfo{journal}{Rev. Mod. Phys.} \textbf{\bibinfo{volume}{80}},
  \bibinfo{pages}{141} (\bibinfo{year}{2008}).

\bibitem[{\citenamefont{Branton et~al.}(2008)\citenamefont{Branton, Deamer,
  Marziali, Bayley, Benner, Butler, {Di Ventra}, Garaj, Hibbs, Huang
  et~al.}}]{Branton2008}
\bibinfo{author}{\bibfnamefont{D.}~\bibnamefont{Branton}},
  \bibinfo{author}{\bibfnamefont{D.~W.} \bibnamefont{Deamer}},
  \bibinfo{author}{\bibfnamefont{A.}~\bibnamefont{Marziali}},
  \bibinfo{author}{\bibfnamefont{H.}~\bibnamefont{Bayley}},
  \bibinfo{author}{\bibfnamefont{S.~A.} \bibnamefont{Benner}},
  \bibinfo{author}{\bibfnamefont{T.}~\bibnamefont{Butler}},
  \bibinfo{author}{\bibfnamefont{M.}~\bibnamefont{{Di Ventra}}},
  \bibinfo{author}{\bibfnamefont{S.}~\bibnamefont{Garaj}},
  \bibinfo{author}{\bibfnamefont{A.}~\bibnamefont{Hibbs}},
  \bibinfo{author}{\bibfnamefont{X.}~\bibnamefont{Huang}},
  \bibnamefont{et~al.}, \bibinfo{journal}{Nat. Biotechnol.}
  \textbf{\bibinfo{volume}{26}}, \bibinfo{pages}{1146} (\bibinfo{year}{2008}).

\bibitem[{\citenamefont{Howorka and Siwy}(2009)}]{Howorka2009}
\bibinfo{author}{\bibfnamefont{S.}~\bibnamefont{Howorka}} \bibnamefont{and}
  \bibinfo{author}{\bibfnamefont{Z.}~\bibnamefont{Siwy}},
  \bibinfo{journal}{Chem. Soc. Rev.} \textbf{\bibinfo{volume}{38}},
  \bibinfo{pages}{2360} (\bibinfo{year}{2009}).

\bibitem[{\citenamefont{Venkatesan and Bashir}(2011)}]{Venkatesan2011}
\bibinfo{author}{\bibfnamefont{B.~M.} \bibnamefont{Venkatesan}}
  \bibnamefont{and} \bibinfo{author}{\bibfnamefont{R.}~\bibnamefont{Bashir}},
  \bibinfo{journal}{Nat. Nanotechnol.} \textbf{\bibinfo{volume}{6}},
  \bibinfo{pages}{615} (\bibinfo{year}{2011}).

\bibitem[{\citenamefont{Storm et~al.}(2001)\citenamefont{Storm, {van Noort},
  {de Vries}, and Dekker}}]{Storm2001}
\bibinfo{author}{\bibfnamefont{A.~J.} \bibnamefont{Storm}},
  \bibinfo{author}{\bibfnamefont{J.}~\bibnamefont{{van Noort}}},
  \bibinfo{author}{\bibfnamefont{S.}~\bibnamefont{{de Vries}}},
  \bibnamefont{and} \bibinfo{author}{\bibfnamefont{C.}~\bibnamefont{Dekker}},
  \bibinfo{journal}{Appl. Phys. Lett.} \textbf{\bibinfo{volume}{79}},
  \bibinfo{pages}{3881} (\bibinfo{year}{2001}).

\bibitem[{\citenamefont{Fologea et~al.}(2005)\citenamefont{Fologea, Uplinger,
  Thomas, McNabb, and Li}}]{Fologea2005a}
\bibinfo{author}{\bibfnamefont{D.}~\bibnamefont{Fologea}},
  \bibinfo{author}{\bibfnamefont{J.}~\bibnamefont{Uplinger}},
  \bibinfo{author}{\bibfnamefont{B.}~\bibnamefont{Thomas}},
  \bibinfo{author}{\bibfnamefont{D.~S.} \bibnamefont{McNabb}},
  \bibnamefont{and} \bibinfo{author}{\bibfnamefont{J.}~\bibnamefont{Li}},
  \bibinfo{journal}{Nano Lett.} \textbf{\bibinfo{volume}{5}},
  \bibinfo{pages}{1734} (\bibinfo{year}{2005}).

\bibitem[{\citenamefont{Dekker}(2007)}]{Dekker2007}
\bibinfo{author}{\bibfnamefont{C.}~\bibnamefont{Dekker}},
  \bibinfo{journal}{Nat. Nanotechnol.} \textbf{\bibinfo{volume}{2}},
  \bibinfo{pages}{209} (\bibinfo{year}{2007}).

\bibitem[{\citenamefont{Tsutsui et~al.}(2012)\citenamefont{Tsutsui, He,
  Furuhashi, Rahong, Taniguchi, and Kawai}}]{Tsutsui2012}
\bibinfo{author}{\bibfnamefont{M.}~\bibnamefont{Tsutsui}},
  \bibinfo{author}{\bibfnamefont{Y.}~\bibnamefont{He}},
  \bibinfo{author}{\bibfnamefont{M.}~\bibnamefont{Furuhashi}},
  \bibinfo{author}{\bibfnamefont{S.}~\bibnamefont{Rahong}},
  \bibinfo{author}{\bibfnamefont{M.}~\bibnamefont{Taniguchi}},
  \bibnamefont{and} \bibinfo{author}{\bibfnamefont{T.}~\bibnamefont{Kawai}},
  \bibinfo{journal}{Sci. Rep.} \textbf{\bibinfo{volume}{2}}, \bibinfo{pages}{1}
  (\bibinfo{year}{2012}).

\bibitem[{\citenamefont{Zwolak and {Di Ventra}}(2005)}]{Zwolak2005}
\bibinfo{author}{\bibfnamefont{M.}~\bibnamefont{Zwolak}} \bibnamefont{and}
  \bibinfo{author}{\bibfnamefont{M.}~\bibnamefont{{Di Ventra}}},
  \bibinfo{journal}{Nano Lett.} \textbf{\bibinfo{volume}{5}},
  \bibinfo{pages}{421} (\bibinfo{year}{2005}).

\bibitem[{\citenamefont{Lagerqvist et~al.}(2006)\citenamefont{Lagerqvist,
  Zwolak, and {Di Ventra}}}]{Lagerqvist2006}
\bibinfo{author}{\bibfnamefont{J.}~\bibnamefont{Lagerqvist}},
  \bibinfo{author}{\bibfnamefont{M.}~\bibnamefont{Zwolak}}, \bibnamefont{and}
  \bibinfo{author}{\bibfnamefont{M.}~\bibnamefont{{Di Ventra}}},
  \bibinfo{journal}{Nano Lett.} \textbf{\bibinfo{volume}{6}},
  \bibinfo{pages}{779} (\bibinfo{year}{2006}).

\bibitem[{\citenamefont{Lagerqvist et~al.}(2007)\citenamefont{Lagerqvist,
  Zwolak, and {Di Ventra}}}]{Lagerqvist2007}
\bibinfo{author}{\bibfnamefont{J.}~\bibnamefont{Lagerqvist}},
  \bibinfo{author}{\bibfnamefont{M.}~\bibnamefont{Zwolak}}, \bibnamefont{and}
  \bibinfo{author}{\bibfnamefont{M.}~\bibnamefont{{Di Ventra}}},
  \bibinfo{journal}{Biophys. J.} \textbf{\bibinfo{volume}{93}},
  \bibinfo{pages}{2384} (\bibinfo{year}{2007}).

\bibitem[{\citenamefont{Tsutsui
  et~al.}(2008{\natexlab{a}})\citenamefont{Tsutsui, Shoji, Taniguchi, and
  Kawai}}]{Tsutsui2008a}
\bibinfo{author}{\bibfnamefont{M.}~\bibnamefont{Tsutsui}},
  \bibinfo{author}{\bibfnamefont{K.}~\bibnamefont{Shoji}},
  \bibinfo{author}{\bibfnamefont{M.}~\bibnamefont{Taniguchi}},
  \bibnamefont{and} \bibinfo{author}{\bibfnamefont{T.}~\bibnamefont{Kawai}},
  \bibinfo{journal}{Nano Lett.} \textbf{\bibinfo{volume}{8}},
  \bibinfo{pages}{345} (\bibinfo{year}{2008}{\natexlab{a}}).

\bibitem[{\citenamefont{Tsutsui
  et~al.}(2008{\natexlab{b}})\citenamefont{Tsutsui, Taniguchi, and
  Kawai}}]{Tsutsui2008b}
\bibinfo{author}{\bibfnamefont{M.}~\bibnamefont{Tsutsui}},
  \bibinfo{author}{\bibfnamefont{M.}~\bibnamefont{Taniguchi}},
  \bibnamefont{and} \bibinfo{author}{\bibfnamefont{T.}~\bibnamefont{Kawai}},
  \bibinfo{journal}{Appl. Phys. Lett.} \textbf{\bibinfo{volume}{93}},
  \bibinfo{pages}{163115} (\bibinfo{year}{2008}{\natexlab{b}}).

\bibitem[{\citenamefont{Tsutsui et~al.}(2009)\citenamefont{Tsutsui, Taniguchi,
  and Kawai}}]{Tsutsui2009}
\bibinfo{author}{\bibfnamefont{M.}~\bibnamefont{Tsutsui}},
  \bibinfo{author}{\bibfnamefont{M.}~\bibnamefont{Taniguchi}},
  \bibnamefont{and} \bibinfo{author}{\bibfnamefont{T.}~\bibnamefont{Kawai}},
  \bibinfo{journal}{Nano Lett.} \textbf{\bibinfo{volume}{9}},
  \bibinfo{pages}{1659} (\bibinfo{year}{2009}).

\bibitem[{\citenamefont{Tsutsui et~al.}(2010)\citenamefont{Tsutsui, Taniguchi,
  Yokota, and Kawai}}]{Tsutsui2010}
\bibinfo{author}{\bibfnamefont{M.}~\bibnamefont{Tsutsui}},
  \bibinfo{author}{\bibfnamefont{M.}~\bibnamefont{Taniguchi}},
  \bibinfo{author}{\bibfnamefont{K.}~\bibnamefont{Yokota}}, \bibnamefont{and}
  \bibinfo{author}{\bibfnamefont{T.}~\bibnamefont{Kawai}},
  \bibinfo{journal}{Nat. Naotechnol.} \textbf{\bibinfo{volume}{5}},
  \bibinfo{pages}{286} (\bibinfo{year}{2010}).

\bibitem[{\citenamefont{Ohshiro et~al.}(2012)\citenamefont{Ohshiro, Matsubara,
  Tsutsui, Furuhashi, Taniguchi, and Kawai}}]{Ohshiro2012}
\bibinfo{author}{\bibfnamefont{T.}~\bibnamefont{Ohshiro}},
  \bibinfo{author}{\bibfnamefont{K.}~\bibnamefont{Matsubara}},
  \bibinfo{author}{\bibfnamefont{M.}~\bibnamefont{Tsutsui}},
  \bibinfo{author}{\bibfnamefont{M.}~\bibnamefont{Furuhashi}},
  \bibinfo{author}{\bibfnamefont{M.}~\bibnamefont{Taniguchi}},
  \bibnamefont{and} \bibinfo{author}{\bibfnamefont{T.}~\bibnamefont{Kawai}},
  \bibinfo{journal}{Sci. Rep.} \textbf{\bibinfo{volume}{2}}, \bibinfo{pages}{1}
  (\bibinfo{year}{2012}).

\bibitem[{\citenamefont{Simon and Gogotsi}(2008)}]{Simon2008}
\bibinfo{author}{\bibfnamefont{P.}~\bibnamefont{Simon}} \bibnamefont{and}
  \bibinfo{author}{\bibfnamefont{Y.}~\bibnamefont{Gogotsi}},
  \bibinfo{journal}{Nat. Materials} \textbf{\bibinfo{volume}{7}},
  \bibinfo{pages}{845} (\bibinfo{year}{2008}).

\bibitem[{\citenamefont{Rica et~al.}(2012)\citenamefont{Rica, Ziano, Salerno,
  Mantegazza, and Brogioli}}]{Rica2012}
\bibinfo{author}{\bibfnamefont{R.~A.} \bibnamefont{Rica}},
  \bibinfo{author}{\bibfnamefont{R.}~\bibnamefont{Ziano}},
  \bibinfo{author}{\bibfnamefont{D.}~\bibnamefont{Salerno}},
  \bibinfo{author}{\bibfnamefont{F.}~\bibnamefont{Mantegazza}},
  \bibnamefont{and} \bibinfo{author}{\bibfnamefont{D.}~\bibnamefont{Brogioli}},
  \bibinfo{journal}{Phys. Rev. Lett.} \textbf{\bibinfo{volume}{109}},
  \bibinfo{pages}{156103} (\bibinfo{year}{2012}).

\bibitem[{\citenamefont{Wei et~al.}(2012)\citenamefont{Wei, Scherer, Bower,
  Andrew, Ryh{\"a}nen, and Steiner}}]{Wei2012}
\bibinfo{author}{\bibfnamefont{D.}~\bibnamefont{Wei}},
  \bibinfo{author}{\bibfnamefont{M.~R.~J.} \bibnamefont{Scherer}},
  \bibinfo{author}{\bibfnamefont{C.}~\bibnamefont{Bower}},
  \bibinfo{author}{\bibfnamefont{P.}~\bibnamefont{Andrew}},
  \bibinfo{author}{\bibfnamefont{T.}~\bibnamefont{Ryh{\"a}nen}},
  \bibnamefont{and} \bibinfo{author}{\bibfnamefont{U.}~\bibnamefont{Steiner}},
  \bibinfo{journal}{Nano Lett.} \textbf{\bibinfo{volume}{12}},
  \bibinfo{pages}{1857} (\bibinfo{year}{2012}).

\bibitem[{\citenamefont{Kimball}(1940)}]{Kimball1940}
\bibinfo{author}{\bibfnamefont{G.~E.} \bibnamefont{Kimball}},
  \bibinfo{journal}{J. Chem. Phys.} \textbf{\bibinfo{volume}{8}},
  \bibinfo{pages}{199} (\bibinfo{year}{1940}).

\bibitem[{\citenamefont{Hamelin}(1996)}]{Hamelin1996a}
\bibinfo{author}{\bibfnamefont{A.}~\bibnamefont{Hamelin}}, \bibinfo{journal}{J.
  Electroanal. Chem.} \textbf{\bibinfo{volume}{407}}, \bibinfo{pages}{1}
  (\bibinfo{year}{1996}).

\bibitem[{\citenamefont{Schoch et~al.}(2008)\citenamefont{Schoch, Han, and
  Renaud}}]{Schoch2008}
\bibinfo{author}{\bibfnamefont{R.~B.} \bibnamefont{Schoch}},
  \bibinfo{author}{\bibfnamefont{J.}~\bibnamefont{Han}}, \bibnamefont{and}
  \bibinfo{author}{\bibfnamefont{P.}~\bibnamefont{Renaud}},
  \bibinfo{journal}{Rev. Mod. Phys.} \textbf{\bibinfo{volume}{80}},
  \bibinfo{pages}{839} (\bibinfo{year}{2008}).

\bibitem[{\citenamefont{Tabard-Cossa et~al.}(2007)\citenamefont{Tabard-Cossa,
  Trivedi, Wiggin, Jetha, and Marziali}}]{Cossa2007}
\bibinfo{author}{\bibfnamefont{V.}~\bibnamefont{Tabard-Cossa}},
  \bibinfo{author}{\bibfnamefont{D.}~\bibnamefont{Trivedi}},
  \bibinfo{author}{\bibfnamefont{M.}~\bibnamefont{Wiggin}},
  \bibinfo{author}{\bibfnamefont{N.~N.} \bibnamefont{Jetha}}, \bibnamefont{and}
  \bibinfo{author}{\bibfnamefont{A.}~\bibnamefont{Marziali}},
  \bibinfo{journal}{Nanotechnol.} \textbf{\bibinfo{volume}{18}},
  \bibinfo{pages}{305505} (\bibinfo{year}{2007}).

\bibitem[{\citenamefont{Smeets et~al.}(2008)\citenamefont{Smeets, Keyser,
  Dekker, and Dekker}}]{Smeets2008}
\bibinfo{author}{\bibfnamefont{R.~M.~M.} \bibnamefont{Smeets}},
  \bibinfo{author}{\bibfnamefont{U.~F.} \bibnamefont{Keyser}},
  \bibinfo{author}{\bibfnamefont{N.~H.} \bibnamefont{Dekker}},
  \bibnamefont{and} \bibinfo{author}{\bibfnamefont{C.}~\bibnamefont{Dekker}},
  \bibinfo{journal}{Proc. Natl. Acad. Sci. U.S.A.}
  \textbf{\bibinfo{volume}{105}}, \bibinfo{pages}{417} (\bibinfo{year}{2008}).

\bibitem[{\citenamefont{Venkatesan et~al.}(2012)\citenamefont{Venkatesan,
  Estrada, Banerjee, Jin, Dorgan, Bae, Aluru, Pop, and
  Bashir}}]{Venkatesan2012}
\bibinfo{author}{\bibfnamefont{B.~M.} \bibnamefont{Venkatesan}},
  \bibinfo{author}{\bibfnamefont{D.}~\bibnamefont{Estrada}},
  \bibinfo{author}{\bibfnamefont{S.}~\bibnamefont{Banerjee}},
  \bibinfo{author}{\bibfnamefont{X.}~\bibnamefont{Jin}},
  \bibinfo{author}{\bibfnamefont{V.~E.} \bibnamefont{Dorgan}},
  \bibinfo{author}{\bibfnamefont{M.-H.} \bibnamefont{Bae}},
  \bibinfo{author}{\bibfnamefont{N.~R.} \bibnamefont{Aluru}},
  \bibinfo{author}{\bibfnamefont{E.}~\bibnamefont{Pop}}, \bibnamefont{and}
  \bibinfo{author}{\bibfnamefont{R.}~\bibnamefont{Bashir}},
  \bibinfo{journal}{ACS Nano} \textbf{\bibinfo{volume}{6}},
  \bibinfo{pages}{441} (\bibinfo{year}{2012}).

\bibitem[{\citenamefont{Press et~al.}(1992)\citenamefont{Press, Teukolsky,
  Vetterling, and Flannery}}]{BM}
\bibinfo{author}{\bibfnamefont{W.~H.} \bibnamefont{Press}},
  \bibinfo{author}{\bibfnamefont{S.~A.} \bibnamefont{Teukolsky}},
  \bibinfo{author}{\bibfnamefont{W.~T.} \bibnamefont{Vetterling}},
  \bibnamefont{and} \bibinfo{author}{\bibfnamefont{B.~P.}
  \bibnamefont{Flannery}}, \emph{\bibinfo{title}{Numerical Recipes in Fortran
  77, 2nd Ed.}} (\bibinfo{publisher}{Cambridge University Press},
  \bibinfo{year}{1992}), pp. \bibinfo{pages}{279--280}.

\bibitem[{\citenamefont{Kharkats}(1979)}]{Kharkats1979}
\bibinfo{author}{\bibfnamefont{Y.~I.} \bibnamefont{Kharkats}},
  \bibinfo{journal}{J. Electroanal. Chem.} \textbf{\bibinfo{volume}{105}},
  \bibinfo{pages}{97} (\bibinfo{year}{1979}).

\bibitem[{\citenamefont{Vidulich et~al.}(1967)\citenamefont{Vidulich, Evans,
  and Kay}}]{Vidulich1967}
\bibinfo{author}{\bibfnamefont{G.~A.} \bibnamefont{Vidulich}},
  \bibinfo{author}{\bibfnamefont{D.~F.} \bibnamefont{Evans}}, \bibnamefont{and}
  \bibinfo{author}{\bibfnamefont{R.~L.} \bibnamefont{Kay}},
  \bibinfo{journal}{J. Phys. Chem.} \textbf{\bibinfo{volume}{71}},
  \bibinfo{pages}{656} (\bibinfo{year}{1967}).

\end{thebibliography}

\end{document}